\documentclass[10pt]{article}
\usepackage{graphicx}
\usepackage{amsmath,amssymb,verbatim,amsthm}
\newtheorem{theorem}{Theorem}[section]
\newtheorem{lemma}[theorem]{Lemma} 

\newtheorem{corollary}[theorem]{Corollary}

\newtheorem{remarks}[theorem]{Remarks}

\newenvironment{prf}{\vspace{1ex}\begin{proof}[\bf Proof]}{\end{proof}}
\numberwithin{equation}{section}

\newcommand{\OP}{O_{\PP}}
\newcommand{\OoP}{O_{\PP_{b_0}}}

\newcommand{\EE}{\mathbb{E}}
\newcommand{\eps}{\varepsilon}
\newcommand{\RR}{\mathbb{R}}
\newcommand{\NN}{\mathbb{N}}
\newcommand{\ZZ}{\mathbb{Z}}
\newcommand{\GG}{\mathbb{G}}
\newcommand{\PP}{\mathbb{P}}

\newcommand{\bbT}{\mathbb{T}}
\newcommand{\lle}{\lesssim}
\newcommand{\bbE}{\mathbb{E}}
\newcommand{\bbP}{\mathbb{P}}
\newcommand{\bbR}{\mathbb{R}}

\newcommand{\bbN}{\mathbb{N}}

\newcommand{\qv}[1]{\left<  #1  \right>}
\newcommand{\FF}{\mathcal{F}}
\newcommand{\given}{\,|\,}

\newcommand{\cC}{\mathcal{C}}

\newcommand{\ito}{It\={o}}

\title{\bf Posterior Consistency via Precision Operators
 for Bayesian Nonparametric Drift Estimation in SDEs}

\author{
Y. Pokern\thanks{Department of Statistical Science, University College London. 
Email: y.pokern@ucl.ac.uk},\,\,
A.M. Stuart\thanks{
  Department of Mathematics, University of Warwick, Coventry CV4 7AL, England. Email: a.m.stuart@warwick.ac.uk 
  },\,\,
J.H. van Zanten\thanks{Korteweg-de Vries Institute for Mathematics, University of Amsterdam,  
The Netherlands. Email: j.h.vanzanten@uva.nl}
}

\begin{document}
\maketitle
%=======================================================================

\begin{center}
{\em Accepted for Publication in Stochastic Processes and their Applications}
\end{center}

\bigskip

\begin{abstract}
We study a Bayesian approach to nonparametric estimation of the periodic drift 
function of a one-dimensional diffusion from continuous-time data. 
Rewriting the likelihood in terms of local time of the process, 
and specifying a Gaussian prior 
with precision operator of differential form,
we show that the posterior is also Gaussian with 
precision operator also of differential form.
The resulting expressions are explicit and lead
to algorithms which are readily implementable.  
Using new functional limit theorems for the local 
time of diffusions on the circle, we bound the rate at which the posterior 
contracts around the true drift function. 
%
%The original abstract did not comply with the 100 word limit imposed by SPA
%
%We study a Bayesian approach to nonparametric estimation of the periodic drift function of a 
%one-dimensional diffusion from continuous-time data. 
%We rewrite the likelihood in terms of Riemann
%integrals, by introducing the local time of the process,
%and specify a centered Gaussian prior on the drift
%with a precision operator that is of differential form. 
%It is proved that this is a conjugate prior for the likelihood
%in the sense that the posterior is also Gaussian. 
%We give an explicit expression for 
%the posterior precision operator, also of differential form,
%and show that the posterior mean is the solution of a 
%differential equation requiring inversion of the posterior
%precision operator for its solution. 
%Moreover, we bound the rate at which the posterior contracts around
%the true drift function. Our formulation of the estimation
%problem leads to algorithms which are readily implementable
%and analyzed using ideas from the numerical analysis of
%differential equations.
%The central results proved here require tools from 
%the analysis of differential equations, together with 
%new  functional limit theorems for the local 
%time of diffusions on the circle. 
\end{abstract}
%
%=======================================================================
\section{Introduction} 
Diffusion processes are routinely used as statistical models
for a large variety of phenomena including molecular dynamics,
econometrics and climate dynamics 
(see for instance \cite{SchlickBook}, \cite{Karatzas2} 
and \cite{Imkeller}). 
Such a process can be specified via the drift
and diffusion functions of a stochastic differential equation
driven by a Brownian motion $W$.
Even in one dimension, this class of processes attracts great applied
interest. In this case, provided the diffusion function $\sigma$ is known and
under mild additional assumptions, one can  transform the process 
such that the diffusion function is constant:
\begin{align}\label{eq:preDiff} 
 dX_t &= b(X_t) dt + dW_t.
\end{align}
This is the form we consider here.

%
%We refer to finding a 
%solution $\{X_t\}_{t=0}^T$ of \eqref{eq:preDiff} given the drift $b$ as the 
%{\em forward} problem which is accessible at textbook level (see e.g.
%\cite{Durrett96, Oksendal00, Karatzas}) analytically in 
%continuous time. Its computational solution in discrete time is summarized in
%\cite{KloedenPlaten} and under additional regularity assumptions on $b$
%the errors usually associated with discretization can even be 
%avoided altogether using exact simulation, see \cite{BesPapRob06}.

We are interested in the statistical problem of recovering the drift function $b$
given an observed path of the diffusion, $\{X_t\}_{t \in [0,T]}$,
which is a solution of \eqref{eq:preDiff}.
Whenever application-driven insight into the form of the drift $b$
is available, one can attempt to exploit this by postulating a parametric model for $b$, indexed by 
 some finite-dimensional parameter $\theta \in \Theta \subset \bbR^d$. 
The statistical problem then reduces to estimating the parameter
 $\theta$, see e.g. \cite{Kutoyants04} 
for an overview of this well-researched area.
In other cases however, one has to resort to nonparametric methods for making inference 
on the function $b$.
Several such methods have been proposed in the literature. 
An incomplete list include kernel methods 
(e.g.\ \cite{Banon}, \cite{Kutoyants04},  \cite{Zanten01}), 
penalized likelihood methods (e.g.\ \cite{Comte}), and spectral approaches  \cite{BandiPhillips}.

In this paper we investigate  recently developed Bayesian methodology for estimating the 
drift function of a diffusion based on continuous-time observations $X^T = \{X_t\}_{t \in [0,T]}$. 
We consider a periodic set-up, which essentially means that we observe a diffusion on the circle.
This is motivated by applications, for instance in molecular dynamics or neuroscience, in which the data consists of recordings of angles, 
cf.\ e.g.\ \cite{Yvo}, \cite{Hindriks} or \cite{PapPokRobStu11}.
We will consider Gaussian prior measures for the periodic drift function $b$ whose 
inverse covariance operators are chosen from a family
of even order differential operators. 
Recent applied work has shown that this is computationally attractive, since numerical methods 
for differential equations can be used for posterior sampling. 
Specifically, for the prior distributions we consider in this paper we will derive a weakly formulated
 differential equation for the posterior mean. Existing numerical methods can be 
 used to solve this equation, allowing for posterior inference without the need to 
 resort to Markov chain Monte Carlo methods. 
This numerical approach, including
 algorithms to accommodate both continuously and discretely observed data,
is detailed in the paper \cite{PapPokRobStu11}.

In Section \ref{sec:Obs} we precisely state the
inference problem of interest, and describe the
properties of the family of Gaussian priors that we adopt.
We postulate a prior precision operator of the form 
\[
 \cC_0^{-1} = \eta\left( \left(-\Delta\right)^p + \kappa I \right), 
\]
where $\Delta$ is the one-dimensional Laplacian, $p$ is an
integer and $\eta, \kappa$ are real and positive hyper-parameters.Working with prior precision operators has numerous
computational advantages and a central goal of this work is
to develop statistical tools of analysis, in particular
for posterior consistency studies, which are well-adapted to
this setting. The work of \cite{ALS12} developed tools
of analysis which do this in the context of linear inverse
problems with small observational noise, and we adapt
the techniques developed there to our setting.

An appealing aspect of choosing a Gaussian prior on the drift function $b$ is conjugacy, 
in the sense that the posterior is Gaussian as well. Since the log-likelihood is quadratic in $b$ 
(Girsanov's theorem)  this is not unexpected. 
Formally the posterior can be computed by ``completing the square''. 
We note however that for our model, if  $b$ is distributed according to a Gaussian prior
$\Pi$ and given $b$, the data $X^T$ are generated by (\ref{eq:preDiff}), 
the joint distribution of $b$ and $X^T$ is obviously {not} Gaussian in general.
As a result, deriving Gaussianity of the posterior in this infinite-dimensional setting is not
entirely straightforward.  
Section \ref{sec:Pos} is devoted to showing that,
for the priors that we consider, the posterior, i.e.\ the conditional distribution 
of $b$ given $X^T$, is indeed Gaussian. 
After a formal derivation of the associated posterior in 
Subsection \ref{ssec: formal} we 
rigorously prove in Theorem \ref{thm: post} below that the associated posterior is Gaussian 
and obtain the posterior mean and covariance structure.  
The posterior precision operator is again a differential operator, involving the local time 
of the diffusion, and the posterior mean is characterized as the unique weak solution of a 
$2p$-th order differential equation. 
In Subsection \ref{ssec: pen} we outline how
%, as in the finite dimensional
%case, 
our Bayesian approach with Gaussian prior can be viewed
as a penalized least-squares estimator, 
where the $p$th order Sobolev norm of $b$ is penalized and the hyper-parameters $\eta$ and 
$\kappa$ quantify the degree of penalization. In the inverse problem
literature this connection is known as Tikhonov-regularisation.

In Bayesian nonparametrics it is well known that careless constructions of priors can lead 
to inconsistent procedures and sub-optimal convergence rates 
(e.g.\ \cite{diaconis}, \cite{castillo}). 
Consistency or rate of convergence results are often obtained using 
general results that are available  
for various types of statistical models and that give sufficient conditions in terms of metric entropy and 
prior mass assumptions. 
 See, for instance, \cite{Ghosh}, \cite{GGvdV}, \cite{GvdV}, 
\cite{MVZ}, and the references therein.
In this paper however we use 
the explicit description of the posterior distribution, which allows us  to take a rather direct approach 
to studying the asymptotic behaviour of our procedure. In particular, we  avoid entropy or prior mass considerations.

Since the posterior involves a periodic version of the local time of the process $X$, 
the asymptotic properties of the local time play a key role in this investigation. 
In the present setting the existing asymptotic theory for the 
local time of ergodic diffusions
(cf.\ e.g.\ \cite{Harry1}, \cite{AadHarry}) can not be used however, 
since we do not assume ergodicity but instead rely on the periodicity of the drift function $b$ 
to accumulate information as $T \to \infty$. As a consequence, the existing posterior rate of convergence results 
for ergodic diffusion models of \cite{PZ} do not apply.
Existing limit theorems for diffusions with periodic coefficients (e.g.\ \cite{Bha}, \cite{Saf}, \cite{Bol79}) also do not 
suffice for our purpose.
In Section \ref{sec:local} we therefore present new limit theorems for the local time of 
diffusions on the circle. 
These can be seen as extending and complementing the work of Bolthausen \cite{Bol79}, 
who proved a uniform central limit theorem 
for the local time of  Brownian motion on the circle (the case $b\equiv 0$ in (\ref{eq:preDiff})). 
For our purposes we need  asymptotic tightness of the properly normalized local time 
in certain  Sobolev spaces however, and we need the result not just for Brownian motion, but 
for general periodic, zero-mean drift functions $b$.

Having these technical tools in place we use them in combination with methods 
from the analysis of differential equations 
in Section \ref{sec:rate} to 
obtain a rate of contraction result for the posterior distribution. 
The result states
that when the true drift function $b$ is periodic and $p$-regular in the Sobolev sense, 
then the posterior contracts around $b$ at a rate that is essentially $T^{-(p-1/2)/(2p)}$ 
as $T \to \infty$ (with respect to the $L^2$-norm). In particular, we have posterior consistency. 

In the concluding section we discuss several possibilities for further refinements and extensions
of the present work.

\section{Observation model and prior distribution}\label{s:problem} 
\label{sec:Obs}

In this section we first introduce the diffusion process under study, fixing
notation and describing how we exploit periodicity; see
Subsection \ref{ssec:diff}. 
In Subsection \ref{ss:priorHeu} we introduce the  prior we place 
on the drift function of the diffusion,  specifying 
the prior precision operator and collecting basic properties.

\subsection{The diffusion}
\label{ssec:diff}

Consider the stochastic differential equation (SDE)
\begin{align} \label{eq:diff}
dX_t = b(X_t)\,dt + dW_t,\quad X_0=0, 
\end{align}
where $W$ is standard Brownian motion and $b: \bbR \to \bbR$ is a 
continuously differentiable, 
$1$-periodic drift function
with zero mean, i.e.\  $b(x+k)=b(x)$ for all $x\in \mathbb{R}$ and $k\in\mathbb{Z}$
and  $\int_0^1b(x) dx=0$. We let $\mathbb{T}$ denote the
circle $[0,1)$ so that we can also write $b:\mathbb{T} \rightarrow \mathbb{R}$
and we summarize the assumptions on $b$ by writing $b \in \dot{C}^1(\bbT)$,
the dot denoting mean zero.

We assume the mean zero property of $b$ for technical reasons. From the perspective 
of the statistical problem of nonparametrically estimating the drift $b$ it is not a serious  
restriction. Note that if the diffusion $X$ has a periodic drift function $b$ with mean 
$\bar b= \int_0^1b(x) dx$, then the process $\{X_t - \bar bt\}_
{t \ge 0}$ has the zero-mean drift function 
$b - \bar b$. In practice, the 
mean can be removed in a preliminary step using an auxiliary estimator for $\bar b$.
 The simple estimator $X_T /T$ can be used for instance. It converges in probability to $\bar b$ at 
the rate ${T^{-1/2}}$ as $T \to \infty$ (cf.\ \cite{Bha}, Theorem 3), which is faster than the rates we obtain for the nonparametric problem of estimating the centred drift function.

For every  $b \in \dot{C}^1(\bbT)$ the  SDE (\ref{eq:diff}) has a unique weak solution 
(see e.g. Theorems 6.1.6 and 6.2.1 in \cite{Durrett96}, p. 214). 
For $T > 0$, we denote the law that 
this solution generates on the  canonical path space $C[0,T]$ by $\bbP^T_b$. 
In particular $\bbP^T_0$ is the 
Wiener measure on $C[0,T]$. By Girsanov's theorem 
 the laws $\bbP^T_b$, $b \in \dot C^1(\bbT)$, are all equivalent on $C[0,T]$. 
If two measurable  maps of $X$ are almost surely (a.s.) equal under some $\bbP_{b_0}^T$ 
they are therefore 
a.s.\ equal under any of the laws $\bbP_b^T$, and we will simply write that they are equal a.s..

We drop the superscript $T$ 
and denote the sample path of \eqref{eq:diff}
by $X \in C[0,T]$. 
The Radon-Nikodym derivative of $\bbP_b^T$  
relative to the Wiener measure satisfies
\[
\label{eq:G1}
\frac{d\bbP^T_b}{d\bbP^T_0}(X)=
\exp\Bigl(- \frac12 \int_0^T b^2(X_t)\, dt + \int_0^T  b\bigl(X_t\bigr)\,dX_t\Bigr)
\]
almost surely, by Girsanov's theorem (e.g.\ \cite{Liptser}). 
Observe that by  \ito's formula the likelihood can be rewritten as 
\begin{align*}
\frac{d\bbP^T_b}{d\bbP^T_0}(X)=\exp \Bigl( 
- \Phi_T(b;X)\Bigr)
\end{align*}
a.s., where 
\begin{align}\label{eq:phiPath}
\Phi_T(b;X) = \frac12 \int_0^T \left(
b^2(X_t) + b'(X_t) \right) dt + B(X_0)-B(X_T)
\end{align}
and $B'=b$. Note that $B$ is also $1$-periodic, since $b$ has average zero.

It will be convenient to write the integrals in the expression 
for $\Phi_T$ in terms of the local time of the process $X$. 
Let $(L_t(x; X): t \ge 0, x \in \bbR)$ be the  semi-martingale local time of $X$, so that 
\begin{equation}\label{eq: occs}
 \int_{-\infty}^\infty f(x)L_T(x; X)\,dx = \int_0^T f(X_s)\,ds
\end{equation}
 holds a.s.\ for any bounded, measurable $f:\bbR \rightarrow \bbR$.
Defining also the random variables $\chi_T(x; X)$ by 
\[
 \chi_T(x; X) = 
 \begin{cases}
1 & \text{if} \ X_0<x<X_T, \\
-1 & \text{if} \ X_T <x < X_0, \\
0 & \text{otherwise},
\end{cases}
\]
we may then write
\begin{align}\label{eq:phiLocNP}
\Phi_T (b;X) = \frac{1}{2}
\int_\bbR  \Big( L_T(x;X) (b^2(x) +b'(x))-
2 \chi_T(x;X)b(x) \Big)\, dx.
\end{align}

In view of the periodicity of the functions involved it is sensible 
to introduce a periodic version $L^\circ$ of the local time $L$ by defining 
\begin{align*}
 L^\circ_T(x; X) &= \sum_{k\in \mathbb{Z}} L_T(x+k; X) 
\end{align*}
for $x \in \bbT$. Note that for every $T > 0 $,
the random function $x \mapsto L_T(x; X)$ 
a.s.\ is a continuous function with support included in the compact interval 
$[\min_{t \le T} X_t, \max_{t \le T} X_t]$. 
Hence the infinite sum can actually be restricted to the finitely many integers
in the interval $[\min_{t \le T} X_t -1, \max_{t \le T} X_t +1]$.
Hence the sum is well defined and $x \mapsto  L^\circ_T(x; X)$ is a continuous random function on $\bbT$.
In particular, we have that the norms $\|L^\circ_T(\cdot; X)\|_\infty$ 
and $\|L^\circ_T(\cdot; X)\|_{L^2}$ are a.s.\ finite.

It follows from (\ref{eq: occs}) that 
for any $1$-periodic, bounded, measurable function $f$ and $T \ge 0$,  
\begin{equation}\label{eq: occp1}
\int_0^T f(X_u)\,du = \int_0^1 f(x)L^\circ_T(x; X)\,dx.
\end{equation}
Exploiting the periodicity of $b$ and $B$ and introducing the corresponding
periodized version $\chi_T^\circ(\cdot;X)$ 
of $\chi_T(\cdot;X)$, we can then rewrite \eqref{eq:phiLocNP} as
\begin{align}\label{eq:phiLoc}
 \Phi_T (b;X) = \frac{1}{2}
\int_0^1 \Big(L_T^\circ(x;X) (b^2(x) +b'(x))-
2 \chi_T^\circ(x;X)b(x) \Big)\, dx.
\end{align}

Summarizing, we have the following lemma. 

\begin{lemma}
For every  $b \in \dot{C}^1(\bbT)$ and $T > 0$ the law $\bbP^T_b$ is equivalent to $\bbP^T_0$ on $C[0,T]$ and 
\begin{align*}
\frac{d\bbP^T_b}{d\bbP^T_0}(X)=\exp \Bigl( 
- \Phi_T(b;X)\Bigr),
\end{align*}
a.s., where $\Phi_T$ is given by (\ref{eq:phiLoc}).  
\end{lemma}

\subsection{The prior }\label{ss:priorHeu}
%and Heuristics for the Posterior

We will assume that we observe a solution of the SDE (\ref{eq:diff}) up to time $T > 0$, 
for some $b \in \dot C^1(\bbT)$.  
To  make inference on $b$ we endow it with a  centred Gaussian prior $\Pi$. 
We will view the prior as a centred Gaussian measure on  $L^2(\bbT)$ and 
define it through its covariance operator $\mathcal{C}_0$, or, rather, 
through its precision operator $\cC_0^{-1}$. 
Specifically, we fix hyper-parameters $\eta, \kappa > 0$ and 
$p \in \{2, 3, \ldots\}$ and consider the operator $\cC_0$ with densely defined inverse 
\begin{align}
 \cC_0^{-1} = \eta\left( \left(-\Delta\right)^p + \kappa I \right) \label{eq:priorCov},
\end{align}
where $\Delta$ denotes the one-dimensional
Laplacian, $I$ is the identity and the domain of $\cC_0^{-1}$ is given by 
$D(\cC_0^{-1})=\dot{H}^{2p}(\mathbb{T})$,  
the space of mean-zero functions in the Sobolev space $H^{2p}(\bbT)$ of 
functions in $L^2(\bbT)$ with $2p$ square integrable weak derivatives. 
 
To see that $\cC_0$ is indeed a valid covariance operator and hence the prior is well defined, 
consider the orthonormal  basis $\phi_k$ of  
$\dot{L}^2(\bbT)$, which is by definition the space of mean-zero functions  in $L^2(\bbT)$, 
given by
\begin{align*}
\phi_{2k}(x) & =\sqrt{2} \cos(2\pi kx),\\
\phi_{2k-1}(x) &=\sqrt{2} \sin(2\pi kx), 
\end{align*}
for $k \in \mathbb{N}$. 
The functions $\phi_k$ belong to the domain $\dot{H}^{2p}(\mathbb{T})$ of the operator 
(\ref{eq:priorCov}) and 
\begin{align*}
\cC_0^{-1}\phi_{2k} &=\bigl(\eta (4\pi^2 k^2)^{p}+\eta \kappa\bigr)
\phi_{2k},\\
\cC_0^{-1}\phi_{2k-1} & =\bigl(\eta(4\pi^2 k^2)^{p}+ \eta\kappa\bigr)
\phi_{2k-1},
\end{align*}
for $k \in \bbN$. 
It follows that $\cC_0$ is the operator on $\dot L^2(\bbT)$ 
which is diagonalized by the basis $\phi_k$, with eigenvalues  
\begin{equation}
\lambda_k = \Big(\eta \Big( 4\pi^2 \Big\lceil \frac{k}{2} \Big\rceil^2 \Big)^{p}
+\eta \kappa\Big)^{-1}. 
\end{equation}
Thus $\cC_0$ is positive definite,  symmetric, and trace-class 
and hence a covariance operator on $\dot L^2(\mathbb{T})$.
(It extends to a covariance operator on the whole space $L^2(\bbT)$ by setting $\cC_01= 0$.)

The integer $p$ in (\ref{eq:priorCov}) controls the regularity of the prior $\Pi$ and we assume $p \ge 2$ to ensure
that the drift is $C^1$ (see Lemma \ref{lem: prior} below). The
parameter $\eta>0$ sets an overall scale
for the precision. The parameter $\kappa$ allows us to shift 
the precisions in every mode by a uniform amount. We employ
$\kappa>0$ as it simplifies some of the analysis, but $\kappa=0$
could be included in the analysis with further work.
Likewise we have assumed a mean zero prior, but extensions
to include a mean could be made.

The preceding calculations show that the prior $\Pi$ is the law of the centred 
Gaussian process $V = \{V_x\}_{x \in \bbT}$ defined by 
\begin{equation}\label{eq: karhunen}
V_x = \sum_{k \in \bbN} \sqrt{\lambda_k} \phi_k(x) Z_k, 
\end{equation}
for $Z_1, Z_2, \ldots$ independent, standard Gaussian random variables. 
Using this series representation a number of basic properties of the 
prior can easily be derived. 

\begin{lemma}~\label{lem: prior}
\begin{enumerate}
\item[(i)]
There exists a version of $V$ which a.s.\ has sample path that are H\"older continuous 
of order $\alpha$, for every $\alpha < p -1/2$. 
\item[(ii)]
The reproducing kernel Hilbert space of $V$ is the Sobolev space $\dot H^{p}(\bbT)$.
\item[(iii)]
The $L^2$-support of $\Pi$ is $\dot L^2(\bbT)$.
\end{enumerate}
\end{lemma}

\begin{prf}
For the first statement, 
note that $\sqrt{\lambda_k} \sim k^{-p}$ asymptotically. 
Using also the  differential relations between the basis functions  $\phi_k$ 
it is straightforward to see that the process $V$ has $p-1 \ge 1$ weak 
derivatives in the $L^2$-sense. Moreover, using Kolmogorov's classical continuity theorem 
it can be shown that this $(p-1)$st derivative 
has a version with sample paths that are H\"older continuous of order $\gamma$ 
for every $\gamma < 1/2$. 
Combining this we see that $V$ has a version with $\alpha$-H\"older sample paths, 
for every $\alpha < p-1/2$. 
In particular, it holds that all the mass of the prior  $\Pi$ is concentrated on  $\dot C^1(\bbT)$. 

The Karhunen-Lo\`eve expansion (\ref{eq: karhunen}) 
shows that the reproducing kernel Hilbert space (RKHS) of the prior is 
given by $\mathbb{H} = \{\sum_{k \ge 1} c_k \phi_k : \sum c^2_k/\lambda_k < \infty\}$
(see for instance \cite{RKHS}, Theorem 4.1).
Since $1/\lambda_k \sim k^{2p}$ this implies that $\mathbb{H} = \dot H^{p}(\bbT)$, 
proving the second statement. 

The final statement follows  from the second one, since the $L^2$-support is 
the $L^2$-closure of $\mathbb{H}$ (\cite{RKHS}, Lemma 5.1).
\end{prf}

Note in particular that the lemma shows that 
we can view the prior $\Pi$ as a Gaussian measure on any of the separable Banach spaces 
$L^2(\bbT)$, $C(\bbT)$, $C^k(\bbT)$ or $H^k(\bbT)$,  for $k \le p-1$. 

\section{Posterior distribution}
\label{sec:Pos}

\subsection{Bayes' formula}
\label{sec: bayes}

We recall that $X$ denotes the path $\{X_t\}_{t \in [0,T]}.$
If we endow  $C^1(\bbT)$ with its H\"older norm and $C[0,T]$ with the uniform norm, then 
expression (\ref{eq:phiPath}) shows that the negative 
log-likelihood $(b, x) \mapsto \Phi_T(b;x)$ (has a version that) 
is Borel-measurable as a map from $C^1(\bbT) \times C[0,T] \to \bbR$.  
Since we can view $\Pi$ as a measure on $C^1(\bbT)$,  
it follows that we have a well-defined Borel measure 
$\Pi(db)\bbP_b^T(dx)$ on $C^1(\bbT) \times C[0,T]$, 
which is the joint law of $b$ and $X$ in the Bayesian set-up 
\[
b \sim \Pi, \qquad X \given b \sim (\ref{eq:diff}).
\]
The posterior distribution, i.e.\ the  conditional distribution of $b$ given $X$, 
is then well-defined as well and given by 
\begin{equation}\label{eq: z}
\begin{split}
\Pi(B \given X) & = \frac1 Z \int_B \exp \left(-\Phi_T(b;X)\right)\Pi(db), \\
\qquad Z & = \int_{C^1(\bbT)} \exp \left(-\Phi_T(b;X)\right)\Pi(db), 
\end{split}
\end{equation}

\begin{lemma}
The random Borel measure $B \mapsto \Pi(B \given X)$ on $C^1(\bbT)$ given 
by (\ref{eq: z}) is a.s.\ well defined.  
\end{lemma}

\begin{prf}
By Lemma 5.3 of \cite{HSV05}, the posterior is well-defined if 
 $Z >0$ a.s..
To see that the latter condition is fulfilled, observe that 
\[
|\Phi_T(b; X)| \lle (1+\|L^\circ_T(\cdot; X)\|_\infty)(\|b\|^2_\infty + \|b'\|_\infty).
\]
Since $\Pi$ is 
a centred Gaussian distribution 
on the separable Banach space  $C^1(\bbT)$, endowed with its H\"older norm, we have 
$\Pi(b: \|b\|_\infty + \|b'\|_\infty < \infty) = 1$. 
Together this gives the a.s.\ positivity of $Z$, 
since $\|L^\circ_T(\cdot; X)\|_\infty < \infty$ a.s.
(Here, and elsewhere, $a \lle b$ means that $a$ is less than an irrelevant constant times $b$.) 
\end{prf}

We have  now defined the posterior as a  measure on $C^1(\bbT)$, 
but since the prior is in fact 
a probability measure on $C^\alpha(\bbT)$ for every $\alpha < p-1/2$ (see the preceding section), 
it is a Borel measure on these H\"older spaces as well.
We can of course also view it as a measure on $C(\bbT)$ or $L^2(\bbT)$.

\subsection{Formal computation of the posterior}
\label{ssec: formal}

The next goal is to characterize the posterior. We proceed first 
strictly formally and non-rigorously. Very loosely speaking, 
we have that  the prior $\Pi$ has a ``density'' proportional to 
\begin{equation}\label{eq: density}
b \mapsto \exp\Big(-\frac12 {\int_0^1 b(x) \cC^{-1}_0b(x)\,dx}\Big)
\end{equation}
and the negative log-likelihood also has a quadratic form, given by (\ref{eq:phiLoc}). 
This suggests that the posterior is again Gaussian. 
Formally completing the square gives the relations 
\begin{align}
 \cC_T^{-1} &= \cC_0^{-1}+L^\circ_T(\cdot; X) I \label{eq:covOp},\\
 \cC_T^{-1} \hat b_T &= \frac{1}{2} (L^\circ_T(\cdot; X))' + \chi_T^\circ(\cdot;X) \label{eq:PDE}
\end{align}
for the posterior mean $\hat b_T$ and the posterior precision operator $\cC_T^{-1}$. 

As detailed in the preceding section we assume that the prior covariance operator is 
given by (\ref{eq:priorCov}), with integer $p\geq 2$, $\eta,\kappa >0$, 
$\Delta$ the one-dimensional Laplacian and $D(\cC_0^{-1})=\dot{H}^{2p}(\mathbb{T})$. 
In that case (\ref{eq:covOp}) gives 
\begin{align}
 \cC^{-1}_T = \eta \left(-\Delta\right)^{p} + (\eta\kappa+L^\circ_T(\cdot; X)) I\label{eq:postCov}
\end{align}
and $D(\cC_T^{-1})=\dot{H}^{2p}(\mathbb{T}).$ 
By standard application of the Lax-Milgram theory
(see \cite{Evans}, Section 6.2),  it follows that
the equation $\cC_T^{-1} f = g$ has a unique weak solution in 
$\dot{H}^{p}(\mathbb{T})$ for every $g \in \dot H^{-p}(\bbT)$; 
see \cite{Robinson}, Appendix A, for definition
and properties of the Sobolev spaces $\dot{H}^{2p}(\mathbb{T})$.
From this it follows that $\cC_T$ 
is well defined on all of $\dot H^{-p}(\bbT)$. 
Moreover,  $\cC_T$ is a bounded operator from
$\dot H^{-p}(\bbT)$ into $\dot{H}^{p}(\mathbb{T})$,
since $\cC^{-1}_T$ is coercive.  
If $g \in L^2(\bbT)$ then the weak solution is
more regular and, in fact, lies in $\dot{H}^{2p}(\mathbb{T})$;
see \cite{Evans}, Section 6.3.

The ordinary derivative of local time is not defined,
and indeed is not an element of $L^2(\bbT)$. Thus
we will have interpret (\ref{eq:PDE}) in a weak sense. 
In order to enable
us to do this, in Subsection \ref{ssec: weak} we consider 
the variational formulation of equation (\ref{eq:PDE}).
As as precursor to this, the next subsection is devoted
to observing that the differential equation for the
mean arises 
as the Euler-Lagrange equation for a certain variational problem, 
yielding an interesting connection with penalized least-squares estimation.

\subsection{Connection with penalized least squares}
\label{ssec: pen}

Here we demonstrate the fact
that the  posterior mean $\hat b_T$ given by (\ref{eq:PDE}) 
can be viewed as a penalized least-squares estimator 
in the case $p = 2$. 
Formally, the SDE (\ref{eq:diff}) can be written as 
\[
\dot X_t = b(X_t) + \dot W_t, 
\]
where the dot denotes differentiation with respect to $t$ 
(obviously, the derivatives $\dot X$ and $\dot W$ do not exist in the ordinary sense.) 
This is just a continuous-time version of a standard nonparametric regression model and for 
a drift function $u$, we can view the integral
\[
\int_0^T (\dot X_t - u(X_t))^2\,dt 
\]
as a residual sum of squares. A penalized least-squares procedure consists in adding 
a penalty term to this quantity and minimizing the resulting criterion over $u$. 
Expanding the square in the preceding integral shows that this is equivalent to minimizing 
\[
u \mapsto -\int_0^T u(X_t)\,dX_t + \frac12\int_0^T u^2(X_t)\,dt + P(u), 
\]
over an appropriate space of functions, where $P(u)$ is the penalty. 

If the function $u$ is smooth and periodic, then by It\^o's formula 
and the definitions of $L^\circ$ and $\chi^\circ$, we have, with $U$ a primitive function of $u$,  
\begin{align*}
\int_0^T u(X_t)\,dX_t & = U(X_T) - U(X_0) - \frac12\int_0^T u'(X_t)\,dt \\
& = \int_0^1 u(x)\chi^\circ_T(x; X)\,dx -\frac12\int_0^1 L^\circ_T(x; X)u'(x)\,dx
\end{align*}
and
\[
\int_0^T u^2(X_t)\,dt  = \int_0^1 u^2(x)L^\circ_T(x; X)\,dx.
\]
Hence, if the functions $u$ over which the minimization takes place are smooth enough, 
the criterion can also be written as 
\[
u \mapsto \int_0^1 \Big(\frac12  u^2(x)L^\circ_T(x; X) +\frac12u'(x)L^\circ_T(x; X) 
-u(x)\chi^\circ_T(x; X) \Big)\,dx + P(u). 
\]

Now consider a Sobolev-type penalty term of the form 
\[
P(u) = \frac12 \eta \left( \kappa \int_0^1 (u(x))^2\,dx +  \int_0^1 (u''(x))^2\,dx \right), 
\]
for constants $\eta, \kappa > 0$. Then the objective functional  
$u \mapsto  \Lambda(u;X)$ takes the form  
\[
\Lambda(u;X) =  \int_0^1 \left(\frac12  u^2(\eta\kappa + L^\circ_T(X)) +\frac12u'L^\circ_T( X) 
- u\chi^\circ_T( X)  + \frac12 \eta (u'')^2 \right) \,dx,
\]
where we omitted explicit dependence on $x$ to lighten notation.
To minimize this functional, simply take its variational derivative in the direction
$v$, i.e. compute the limit 
$\lim_{\epsilon\rightarrow 0} \left(\Lambda(u+\epsilon v;X)-\Lambda(u;X)\right)/\epsilon$,
for a smooth test function $v$:
\[
 \frac{\delta \Lambda}{\delta u}(v)= 
\int_0^1 \left( uvL^\circ_T(X)-\frac12 v(L_T^\circ)'(X) - v\chi^\circ_T(X) +\eta v''u''
+\eta \kappa uv \right) \, dx
\]
A further integration by parts (where the boundary terms vanish due to periodicity) now yields
the form
\[
 \frac{\delta \Lambda}{\delta u}(v)= 
\int_0^1 v \left( uL^\circ_T(X)-\frac12 (L_T^\circ)'(X) - \chi^\circ_T(X) +\eta u^{(4)}
+\eta \kappa u \right) \, dx
\]
from which it is evident that equating the variational derivative to zero for all
smooth test functions yields exactly the posterior mean obtained in (\ref{eq:PDE})
for the case $p=2$:
\[
 \eta u^{(4)} + \left(\eta \kappa +L^\circ_T(X)\right)u = \frac12 (L_T^\circ)'(X) +
\chi_T^\circ(X) .
\]
In the context of inverse problems, adding the square of the norm of the underlying
vector space is known as (generalized) Tikhonov regularization, and the connection
to Bayesian inference with a Gaussian prior is well established in general, see
\cite{Tarantola}.
It may be viewed as a natural extension of the
approach of Wahba \cite{Wahba} from regression to
the diffusion process setting. 
The case of regularization through higher order
derivatives in the penalization term $P$ is similar.

\subsection{Weak variational formulation for the posterior mean}
\label{ssec: weak}

In the preceding section we remarked that the RKHS of the Gaussian prior
equals the Sobolev space $\dot H^p(\bbT)$.  
Below we prove that the posterior is a.s.\ a Gaussian measure. Moreover,  since the denominator 
$Z$ in (\ref{eq: z}) is positive a.s., the posterior
is equivalent to the prior.
It follows that the posterior mean $\hat b_T$ is a.s.\ an element of $\dot H^p(\bbT)$. 
By saying it is a weak solution to (\ref{eq:PDE}) we mean that 
it  solves the following weak form of the associated variational principle:
\begin{equation}\label{eq: var}
a(\hat b_T, v; X) = r(v; X) \qquad \text{for every $v \in \dot H^p(\bbT)$},
\end{equation}
where the bilinear form $a(\cdot, \cdot; X): \dot H^p(\bbT)\times \dot H^p(\bbT) \to \bbR$ and 
the linear form $r(\cdot; X): \dot H^p(\bbT) \to \bbR$ are defined by 
\begin{align*}
a(u, v; X) & = \eta\int u^{(p)}(x)  v^{(p)}(x)\,dx + 
\eta\kappa\int u(x)v(x)\,dx + \int u(x)v(x)L^\circ_T(x; X)\,dx, \\
r(v; X) & = -\frac12 \int v'(x)L^\circ_T(x; X)\,dx + \int v(x)\chi^\circ_T(x; X)\,dx.
\end{align*}
The following lemma records the essential properties of $a$ and $r$ and the associated 
variational problem.

\begin{lemma}\label{lem: var}
The following statements hold almost surely:
\begin{enumerate}
\item[(i)]
$a(\cdot, \cdot; X)$ is  bilinear, symmetric, continuous and coercive:
\[
a(v_1, v_2; X) \le  (\eta + \eta\kappa + \|L^\circ_T(\cdot; X)\|_\infty)\|v_1\|_{H^p} \|v_2\|_{H^p}
\]
for $v_1, v_2 \in \dot H^p(\bbT)$ and for some constant $c> 0$, 
$a(v,v; X) \ge c\|v\|^2_{H^p}$ for all $v \in \dot H^p(\bbT)$.
\item[(ii)]
$r(\cdot; X)$ is linear and bounded:
\[
|r(v; X)| \le \Big(\frac12\|L^\circ_T(\cdot; X)\|_{L^2} + 
\|\chi^\circ_T(\cdot; X)\|_{L^2}\Big)\|v\|_{H^p}
\]
for all $v \in \dot H^p(\bbT)$.
\item[(iii)]
There  exists a unique $u \in \dot H^p(\bbT)$ such that $a(u, v; X) = r(v; X)$ 
for all $v \in \dot H^p(\bbT)$.
\end{enumerate}
\end{lemma}

\begin{prf}
(i) Bi-linearity, symmetry and continuity follow straightforwardly from the definition of $a$.  
Coercivity follows easily from the positivity of $\eta$ and
$\kappa$ and the Poincar\'e inequality 
(see \cite{Robinson}, Proposition 5.8.) 
(ii) Again, straightforward. 
(iii) Follows from (i) and (ii) by the Lax-Milgram Lemma, see
\cite{Evans}, Section 6.2. 
\end{prf}

\subsection{Characterization of the posterior}

We can now prove that the posterior is Gaussian and characterize its mean and covariance operator.
Recall that by saying that $\hat b_T$ is a
weak solution of the differential equation 
(\ref{eq:PDE}) we mean 
that it solves the variational problem (\ref{eq: var}).

\begin{theorem}\label{thm: post}
Almost surely,  the posterior $\Pi(\cdot\given X)$ is a 
Gaussian measure on $L^2(\bbT)$. Its covariance operator $\cC_T$ is given by (\ref{eq:postCov})
and its mean $\hat b_T$ is the unique weak solution of (\ref{eq:PDE}). 
\end{theorem}

\begin{prf}
For $n \in \bbN$, let $P_n: L^2(\bbT) \to L^2(\bbT)$ be the orthogonal projection onto 
the linear span $V_n$ of the first $n$ basis functions $\phi_1, \ldots, \phi_n$. 
Let the random measure $\Pi_n(\cdot\given X)$ be given by  
\begin{align*}
\Pi_n(B \given X) & = \frac1 Z_n \int_B \exp \left(-\Phi_T(P_n b;X)\right)\Pi(db), \\
\qquad Z_n & = \int_{C^1(\bbT)} \exp \left(-\Phi_T(P_n b;X)\right)\Pi(db), 
\end{align*}
for Borel sets $B \subset  C^1(\bbT)$. The fact that this random measure 
is well defined can be argued exactly as in Section \ref{sec: bayes}. 

For $b \in \dot C^1(\bbT)$ it holds that $P_nb \to b$ in $H^1(\bbT)$ as $n \to \infty$. 
It is easily seen from (\ref{eq:phiLoc}) that the random map $b \mapsto \Phi_T(b; X)$ is 
a.s.\  $H^1(\bbT)$-continuous. It follows that a.s., 
$b \mapsto \Phi_T(P_nb; X)$ converges point-wise to $\Phi_T(\cdot; X)$ on  $\dot C^1(\bbT)$. 
By Lemma \ref{lem: lower} below, there exists for every $\eps > 0$ a constant $K(\eps)$ such that 
\[
-\Phi_n(b; X) \le \eps\|b\|^2_{H^1} + K(\eps)(1+\|L^\circ_T(\cdot; X)\|^2_{L^2}), 
\]
and hence 
\[
e^{-\Phi_n(P_n b; X)} \le e^{K(\eps)(1+\|L^\circ_T(\cdot; X)\|^2_{L^2})} e^{\eps\|b\|^2_{H^1}} .
\]
Since $\Pi$ can be viewed as a Gaussian measure on $H^1(\bbT)$, Fernique's theorem implies that 
a.s., the right-hand side of the last display is a $\Pi$-integrable function of $b$ for $\eps > 0$ 
small enough (see \cite{Bogachev}, Theorem 2.8.5).  
Hence, by dominated convergence, we can conclude that $Z_n \to Z$ almost surely. 
The same reasoning shows that for every Borel set $B \subset  C^1(\bbT)$, it 
a.s.\ holds that 
\[
\int_B \exp \left(-\Phi_T(P_n b;X)\right)\Pi(db) \to \int_B \exp \left(-\Phi_T(b;X)\right)\Pi(db)
\]
as $n \to \infty$, where we rewrite the integral as an integral over $C^1(\bbT)$ and then
exploit boundedness of the indicator function $\chi_B(\cdot)$ thus introduced into the
integrand.

Hence, we have that with probability $1$, 
the measures $\Pi_n(\cdot\given X)$ converge weakly to the posterior $\Pi(\cdot\given X)$. 
Note that the weak convergence takes place in $C^1(\bbT)$, but then in $L^2(\bbT)$ as well.
Since the measures $\Pi_n(\cdot\given X)$ are easily seen to be Gaussian, the measure 
$\Pi(\cdot\given X)$ must be Gaussian as well.

If we view $\dot L^2(\bbT)$ as the product of $V_n$ and $V_n^\perp$, then by construction  
the measure $\Pi_n(\cdot\given  X)$ is a product of Gaussian measures on $V_n$ and $V_n^\perp$. 
The measure on $V_n$ really has density proportional to 
(\ref{eq: density}), 
relative to the push-forward measure of the Lebesgue measure 
on $\bbR^n$ under the map $(c_1, \ldots, c_n) \mapsto \sum c_k\phi_k$. The formal 
arguments given in Subsection \ref{ssec: formal} can therefore be made rigorous, 
showing that this factor is a Gaussian measure on $V_n$ with covariance 
operator $P_n \cC_T P_n$ and mean $b_n\in V_n$ which solves the variational problem
\[
a(b_n, v; X) = r(v; X)
\]
for every $v \in V_n$. The measure on $V_n^\perp$ has mean zero, 
so $b_n$ is in fact the mean of the whole measure $\Pi_n(\cdot\given  X)$. 
The covariance operator of the measure on $V_n^\perp$ is given by $(I-P_n)\cC_0(I-P_n)$. 

Next we prove that the posterior mean $\hat b_T$ is the weak solution of (\ref{eq:PDE}). 
By Lemma \ref{lem: var} there a.s.\ exists a unique $u \in \dot H^p(\bbT)$ such that 
$a(u, v; X) = r(v; X)$ for all $v \in \dot H^p(\bbT)$. 
Standard Galerkin method arguments show that for the mean of $\Pi_n(\cdot \given X)$ 
we have  $b_n \to u$ in $\dot H^p(\bbT)$. Indeed, let $e_n = u -  b_n$. 
Then we have the orthogonality property $a(e_n, v; X) = 0$ for all $v \in V_n$. 
Using the continuity and coercivity of $a(\cdot, \cdot; X)$, cf.\ Lemma \ref{lem: var}, 
it follows that for $v \in V_n$, 
\begin{align*}
c\|e_n\|^2_{H^p} & \le a(e_n, e_n; X)\\
&  = a(e_n, u-v; X)\\ 
&\le (\eta + \eta\kappa + \|L^\circ_T(\cdot; X)\|_\infty)\|e_n\|_{H^p} \|u-v\|_{H^p}.
\end{align*}
Hence, for every $v \in V_n$ we have
\[
c \|e_n\|_{H^p} \le (\eta + \eta\kappa + \|L^\circ_T(\cdot; X)\|_\infty)\|u-v\|_{H^p}.
\]
By taking $v = P_n u$ we then see that  $b_n \to u$ in $H^p(\bbT)$. 
On the other hand, by the weak convergence found above, $b_n$  converges a.s.\  to 
the posterior mean $\hat b_T$ in $L^2(\bbT)$ (see \cite{Bogachev}, Example 3.8.15).
We conclude that $\hat b_T$ a.s.\ equals the unique weak solution $u$ of (\ref{eq:PDE}), as required.

It remains to show that the covariance operator of the posterior is given by (\ref{eq:postCov}).
Let $\Sigma_n = P_n\cC_TP_n +(I-P_n)\cC_0(I-P_n)$ be the covariance operator 
of $\Pi_n(\cdot\given X)$ and let $\Sigma$ be the covariance operator of the 
posterior $\Pi(\cdot\given X)$. Since the measures converge weakly and are Gaussian, 
we have that for every $f \in L^2(\bbT)$, $\Sigma_n f \to \Sigma f$ in $L^2(\bbT)$ 
(cf.\ Example 3.8.15 of \cite{Bogachev} again). 
On the other hand, for $n > k$ and $g \in \dot L^2(\bbT)$ we have
\begin{align*}
\Big|\qv{g, \Sigma_n\phi_k} - \qv{g, \cC_T\phi_k}_{L^2}\Big| & =
|\qv{g, (P_n-I)\cC_T\phi_k}_{L^2}|\\
&  \le \|(P_n- I)g\|_{L^2} \|\cC_T\phi_k\|_{L^2} \to 0,
\end{align*}
hence $\Sigma_n\phi_k$ converges weakly to $\cC_T\phi_k$. 
It follows that $\Sigma\phi_k = \cC_T\phi_k$ for every $k$ and the proof is complete.
\end{prf}

\begin{lemma}\label{lem: lower}
For every $\eps > 0$ there exists a constant $K(\eps) > 0$ such that 
\[
-\Phi(b; X) \le \eps\|b\|^2_{H^1} + K(\eps)(1+\|L^\circ_T(\cdot; X)\|^2_{L^2}). 
\]
\end{lemma}

\begin{prf}
It follows from  \eqref{eq:phiLoc} that 
\begin{align*}
- \Phi(b;X) \le +\frac{1}{2} \int_0^1 L_T^\circ(x; X) |b'(x)|\,dx + \int_0^1 |b(x)|\, dx.
\end{align*}
Now note that for every $\beta > 0$ and $f, g \in L^2(\bbT)$, it holds that 
$2\qv{f,g} \le \beta \|f\|^2 + \beta^{-1}\|g\|^2$ (``Young's inequality with $\varepsilon$'', with $p=q=2$). 
Applying this to both integrals on the right we get 
\begin{align*}
- \Phi(b;X) & \le \frac{\beta}{4} \|L^\circ_T(\cdot; X)\|^2_{L^2} + 
  \frac1{4\beta}\|b'\|^2_{L^2} + \frac\beta2 + \frac1{2\beta} \|b\|^2_{L^2}\\
& \le  \eps\|b\|^2_{H^1} + \frac1{4\eps} + \frac1{8\eps}\|L^\circ_T(\cdot; X)\|^2_{L^2}, 
\end{align*}
where $\eps = (2\beta)^{-1}$, so $K(\eps)=({4\eps})^{-1}$. 
\end{prf}

\section{Asymptotic behaviour of the local time}
\label{sec:local}

In the next section we will investigate the asymptotic behaviour of the posterior, 
using the characterization provided by Theorem \ref{thm: post}. 
Since \eqref{eq:covOp} and \eqref{eq:PDE} involve the periodic local time 
$L^\circ_T(\cdot; X)$, the asymptotic properties of that random function  play a key role. 

The results  we establish in this section can be seen as complementing and extending 
the work of Bolthausen \cite{Bol79} in which 
it is proved that if $X$ is Brownian motion (i.e.\ $b \equiv 0$ in \eqref{eq:diff}), 
then the random functions 
\[
 \sqrt{T}\Big(\frac{1}{T}L^\circ_T(\cdot; X) - 1\Big)
\]
converge weakly in the space $C(\bbT)$ to a Gaussian random map as $T \to \infty$. 
For our purposes we need a similar result in the 
Sobolev space $H^\alpha(\bbT)$, for $\alpha < 1/2$, and we need the result not just for 
Brownian motion, but for general periodic, 
zero-mean drift functions $b$. In fact asymptotic
tightness, rather than weak convergence,
suffices for our purposes and it is this which we prove. 
In addition, we need the 
associated uniform  law of large numbers which states that 
\[
\frac{1}{T}L^\circ_T(\cdot; X)
\]
converges uniformly as $T \to \infty$. 
Similar statements were obtained for ergodic diffusions in the papers \cite{Harry1} 
and \cite{AadHarry}. In the present periodic setting however, 
completely different arguments are necessary.

Given $b \in \dot C(\bbT)$ we define the probability density $\rho$ on $[0,1]$ by 
\begin{equation}\label{eq: rho}
\rho(x) =  C \exp\Big(2\int_0^x{b(y)}\,dy\Big), \qquad x \in [0,1],
\end{equation}
where $C > 0$ is the normalization constant that ensures that $\rho$ integrates to $1$. 
In the one-dimensional diffusion language, $\rho$ is the restriction to $[0,1]$ 
of the speed density of the diffusion, normalized so that it becomes a probability density. 
Note that since $b$ has mean zero, $\rho$ satisfies 
$\rho(0)=\rho(1)$ and enjoys a natural extension
to a periodic function. 

We use the standard notation that $Y_T = \OP(a_T)$ 
for the family of random variables $\{Y_T\}$ and 
the deterministic family of positive numbers $\{a_T\}$ 
if the family $\{Y_T/a_T\}$ is asymptotically tight 
as $T \to \infty$ with respect to the probability space
underlying the random variables $\{Y_T\}$. 

%We
%will repeatedly use the fact that the
%product of an asymptotically tight sequence with
%any real deterministic sequence which converges to
%zero will converge to zero in probability. 

\begin{theorem}\label{thm: local}~
\begin{enumerate}
\item[(i)]
It almost surely  holds that 
\[
\sup_{x \in \bbT} \Big|\frac{1}{T}L^\circ_T(x; X) - \rho(x)\Big| \to 0
\]
as $T \to \infty$. 
\item[(ii)]
For every $\alpha < 1/2$,  the random maps 
\[
x \mapsto \sqrt{T}\Big(\frac{1}{T}L^\circ_T(x; X) - \rho(x)\Big)
\]
are asymptotically tight in $H^\alpha(\bbT)$ as $T \to \infty$. 
In particular, for every $\alpha < 1/2$, 
\[
\Big\|\frac{1}{T}L^\circ_T(\cdot; X) - \rho\Big\|_{H^\alpha} = \OP\Big(\frac1{\sqrt{T}}\Big)
\]
as $T \to \infty$.
\end{enumerate} 
\end{theorem}

The proof of the theorem is long and therefore deferred to Section \ref{sec:proof} in order to keep the overarching
arguments in this paper, which are aimed to proving posterior 
consistency, to the fore. 
In the following subsection about posterior contraction rates 
we need the fact that $X_T = \OP(\sqrt{T})$ for 
the  diffusions  with periodic drift under consideration. 
This follows for instance from the results of \cite{Bha}, but we can alternatively 
derive it from the preceding theorem. 

\begin{corollary}\label{cor: x}
For every $b \in \dot C(\bbT)$, the weak  solution $X = (X_t: t \ge 0)$ of the SDE (\ref{eq:diff}) 
satisfies  $X_T = \OP(\sqrt{T})$ as $T \to \infty$.
\end{corollary}

\begin{prf}
We have 
\[
X_T =  \int_0^T b(X_s)\,ds + W_T
\]
for a standard Brownian motion $W$. Since $b$ is $1$-periodic, 
the integral can be rewritten in terms of the periodic local time $L^\circ$. 
Moreover, (\ref{eq: rho}) implies that $\rho$ is $1$-periodic as well and $\rho' =2b\rho$, hence 
\[
\int_0^1 b(x)\rho(x)\,dx = \frac1{2}(\rho(1) - \rho(0)) = 0. 
\]
It follows that 
\begin{align*}
|{X_T}| & \le T \Big|\int_0^1 b(x)\Big(\frac{L^\circ_T(x; X)}{T} - \rho(x)\Big)\,dx\Big| + |{W_T}|\\
& \le T \|b\|_{L^2}\Big\|\frac{1}{T}L^\circ_T(\cdot; X) - \rho\Big\|_{L^2} + |W_T|. 
\end{align*}
By the preceding theorem, this is $\OP(\sqrt{T})$.
\end{prf}

Note that statement (i) of Theorem \ref{thm: local} implies that for an integrable, $1$-periodic function $f$, 
we have the strong law of large numbers 
\[
\frac1T \int_0^T f(X_t)\,dt \to \int_0^1f(x)\rho(x)\,dx
\]
a.s.\ as $T \to \infty$. Moreover, if $f$ is also square integrable, statement (ii) implies that 
\[
\frac1T \int_0^T f(X_t)\,dt - \int_0^1f(x)\rho(x)\,dx = \OP\Big(\frac1{\sqrt{T}}\Big).
\]
Result of this type can be found also in \cite{Saf} and are of independent interest. They can for instance 
be useful in the asymptotic analysis of other statistical procedures for the periodic diffusion models we are considering. Uniform Glivenko-Cantelli and Donsker-type
statements could be derived using our approach as well, similar to the results for ergodic diffusions in \cite{Zanten01} and 
\cite{Harry1}. Since this is outside the scope of the present paper however, we do not elaborate on this 
any further here.

\section{Posterior contraction rates}
\label{sec:rate}

In this section we use the characterization of the posterior provided by Theorem \ref{thm: post} 
and the asymptotic behaviour of the local time established in Theorem \ref{thm: local} 
to study the rate at which the posterior contracts around the true drift function, 
which we denote by  $b_0$ to emphasize that the results are 
frequentist in nature. 
In particular $\PP_{b_0}$ denotes the underlying probability measure corresponding to the true drift function
 $b_0$, and the  notation $\OoP$ 
refers to asymptotic tightness
under this measure.

The first theorem concerns the rate of convergence of the posterior mean $\hat b_T$, which, 
by Theorem \ref{thm: post}, is the unique weak solution in $\dot H^p(\bbT)$ of 
the differential equation (\ref{eq:PDE}).

\begin{theorem}\label{thm: mean}
Suppose that the true drift function $b_0 \in \dot H^p(\bbT)$. Then for every $\delta > 0$, 
\[
\|\hat b_T - b_0\|_{L^2} = \OoP(T^{-\frac{p-1/2}{2p} +\delta})
\]
as $T \to \infty$. 
\end{theorem}

\begin{prf}
By Theorem \ref{thm: post} we have (in the weak sense) 
\[
 (\cC_0^{-1}+L^\circ_T(\cdot; X))\hat b_T = \frac{1}{2} (L^\circ_T(\cdot; X))' + 
   \chi_T^\circ(\cdot;X). 
\]
Note that it follows from (\ref{eq: rho}) that  
$\rho$ satisfies $\rho' = 2b_0\rho$ if $b_0$ is the drift function, hence, 
with $\GG_T = \sqrt{T}(L^\circ_T(\cdot; X)/T - \rho)$, 
\begin{align*}
 (\cC_0^{-1}+L^\circ_T(\cdot; X))b_0 & = \frac{1}{2} (L^\circ_T(\cdot; X))' + 
      \chi_T^\circ(\cdot;X)+ \cC_0^{-1}b_0\\
 & \ \ \   + \sqrt{T}\GG_T b_0 - \frac12\sqrt{T}\GG_T' -  \chi_T^\circ(\cdot;X).
 \end{align*}
Subtracting the two equations shows that $e = \hat b_T - b_0$ satisfies (still in the weak sense)
\[
 (\cC_0^{-1}+L^\circ_T(\cdot; X))e =  - \cC_0^{-1}b_0  - \sqrt{T}\GG_T b_0 +
        \frac12\sqrt{T}\GG_T' +  \chi_T^\circ(\cdot;X).
 \]
Since $\rho$ is bounded away from zero (see \eqref{eq: rho}) and 
statement (i) of Theorem \ref{thm: local} says that, almost surely,
$L^\circ_T/T$ converges uniformly on $\bbT$ to $\rho$, it follows 
that there exists a constant $c > 0$ such that 
\[
\inf_{x \in \bbT} \frac{L^\circ_T(x; X)}{T} \ge c
\]
on an event $A_T$ with $\PP_{b_0}(A_T) \to 1$. 
As a consequence, testing the weak differential equation
for the error $e$ with the test function $e$ itself
(``energy method'') yields, 
on the event $A_T$, the inequality
\begin{align*}
& \|\cC_0^{-1/2}e\|^2_{L^2} + cT\|e\|^2_{L^2} \\
& \quad \le |\qv{\cC_0^{-1}b_0, e}| 
 + |\qv{\chi_T^\circ(\cdot;X), e}| + \sqrt{T}|\qv{\GG_T b_0, e}| + \frac12\sqrt{T}|\qv{\GG_T, e'}|\\
& \quad = |\qv{\cC_0^{-\frac12}b_0, \cC_0^{-\frac12}e}| 
 + |\qv{\chi_T^\circ(\cdot;X), e}| + \sqrt{T}|\qv{\GG_T b_0, e}| + \frac12\sqrt{T}|\qv{\GG_T, e'}.
\end{align*}
We now use Young's inequality (\cite{Robinson}, Lemma 1.8)
in the form
$$2\qv{f,g} \le \kappa \|f\|^2_{L^2} + \kappa^{-1}\|g\|^2_{L^2}$$
to estimate the first three terms on the right. 
Choosing appropriate $\kappa$'s and subtracting the 
resulting terms involving $T\|e\|_{L^2}^2$ from both sides we get, still on $A_T$, 
\begin{equation}\label{eq: piet}
\begin{split}
& \|\cC_0^{-1/2}e\|^2_{L^2}  + cT\|e\|^2_{L^2}\\
& \quad \lle \|\cC_0^{-1}b_0\|^2_{L^2} + \frac{\|\chi_T^\circ(\cdot;X)\|^2_{L^2}}{T} +
\|b_0\|^2_\infty \|\GG_T\|^2_{L^2}
+ \frac12\sqrt{T}|\qv{\GG_T, e'}|.
\end{split}
\end{equation}
We note that the 
first three terms on the right are now stochastically bounded: 
the first one is constant, the second is bounded by a $|X_T-X_0|^2/T$, 
which is $\OP(1)$ according to Corollary \ref{cor: x}, 
and the third one is $\OP(1)$ by Theorem \ref{thm: local}. 

For the last term on the right we have, since the norm $\|\cC_0^{-s/(2p)}\cdot\|_{L^2}$ 
is equivalent to the $H^s$-norm and $\cC_0^{{1}/{(2p)}}\frac{\partial}{\partial x}$ is bounded, 
\[
|\qv{\GG_T, e'}| = \Big|\qv{\cC_0^{-\frac{s}{2p}}\GG_T, \cC_0^{\frac{s}{2p}}e'}\Big| 
\lle  \|\GG_T\|_{H^s}\|\cC_0^{\frac{s-1}{2p}} e\|_{L^2}.  
\]
We now use the interpolation inequality given as 
Theorem 13 on p.149 in \cite{DautrayLions},
$$\|A^\iota u \| \leq \|A u \|^\iota \|u \|^{1-\iota},$$
which is valid for $\iota \in (0,1)$ and positive, coercive, self-adjoint
densely defined operators $A$. We take $A=\cC_0^{-\frac12}$ 
and $\iota=({1-s})/{p}$ and combining with what we had above we get \\
\[
\sqrt{T}|\qv{\GG_T, e'}| \lle T^{-\frac{s-1}{2p}} 
     \|\GG_T\|_{H^s}\|\cC_0^{-\frac{1}{2}} e\|^{\frac{1-s}{p}}_{L^2} 
\|\sqrt{T}e\|^{\frac{p+s-1}{p}}_{L^2}.
\]
Using Young's inequality again we have the further bound 
\[
\sqrt{T}|\qv{\GG_T, e'}| \lle \kappa T^{-\frac{s-1}{p}} \|\GG_T\|^2_{H^s} + \kappa^{-1}
\|\cC_0^{-\frac{1}{2}} e\|^2_{L^2} + \kappa^{-1}T\|e\|_{L^2}^2.
\]
If we combine this with (\ref{eq: piet}), choose $\kappa$ large enough and subtract 
$\kappa^{-1}
\|\cC_0^{-\frac{1}{2}} e\|^2_{L^2} + \kappa^{-1}T\|e\|_{L^2}^2$ from both sides of 
the inequality we arrive at the bound 
\[
T\|e\|^2_{L^2}  \le \OP(1) +  T^{-\frac{s-1}{p}} \|\GG_T\|^2_{H^s},
\]
which holds on the event $A_T$. In view of Theorem \ref{thm: local} and 
since $s \in (0,1/2)$ is arbitrary, this completes the proof. 
\end{prf}

Theorem \ref{thm: mean} only concerns the posterior mean, but we can in fact show that 
the whole posterior distribution contracts around the true $b_0$ at the same rate. 
As usual, we say that the posterior contracts around $b_0$ at the rate $\eps_T$ 
(relative to the $L^2$-norm) if for arbitrary positive numbers $M_T \to \infty$, 
\[
\bbE_{b_0}\Pi(b: \|b-b_0\|_{L^2} \ge M_T\eps_T \given X) \to 0 
\]
as $T \to \infty$. This essentially says that for large $T$, the posterior mass is concentrated in 
$L^2$-balls around $b_0$ with a radius of the order $\eps_T$.

\begin{theorem}\label{thm: rate}
Suppose that $b_0 \in \dot H^p(\bbT)$. Then for every $\delta > 0$, 
the posterior contracts around $b_0$ at the rate $T^{-(p-1/2)({2p}) +\delta}$ as $T \to \infty$. 
\end{theorem}

\begin{prf}
Set $\eps_T = T^{-(p-1/2)({2p}) +\delta}$ and consider arbitrary positive numbers $M_T \to \infty$. 
By the triangle inequality, 
\begin{align*}
& \bbE_{b_0}\Pi(b: \|b-b_0\|_{L^2} \ge M_T\eps_T \given X)\\
& \quad \le \bbE_{b_0}\Pi\Big(b: \|b-\hat b_T\|_{L^2} \ge \frac{M_T\eps_T}{2} \given X\Big)
+ \PP_{b_0}\Big(\|\hat b_T-b_0\|_{L^2} \ge \frac{M_T\eps_T}{2}\Big).
\end{align*}
By Theorem \ref{thm: mean} the second term on the right vanishes as $T \to \infty$, hence, 
since the posterior measure of a set is bounded by $1$,  it
suffices to show that $\Pi(b: \|b-\hat b_T\|_{L^2} \ge {M_T\eps_T}/{2} \given X)$
converges to $0$ in $\PP_{b_0}$-probability.
% (recall that since
% $\Pi(b: \|b-b_0\|_{L^2} \ge M_T\eps_T \given X)$ is bounded in modulus by one, convergence
%in probability implies convergence in expectation, see Ex. 7.2.2, p.317 in \cite{Grimmett}).
By Markov's inequality, this quantity is  bounded by 
\[
\frac{4}{M^2_T\eps^2_T} \int \|\hat b_T-b\|^2_{L^2}\Pi(db\given X). 
\]
Since the  integral is equal to the trace of the covariance operator of the centred posterior, it 
suffices to show that $\text{tr}(\cC_T) = O_{\PP_{b_0}}(\eps^2_T)$. 
Since $M_T \to \infty$ this shows that that bound converges
to zero in $\PP_{b_0}$-probability.

As before, let $\lambda_i$ and $\phi_i$ be the eigenvalues and eigenfunctions of 
the prior covariance operator $\cC_0$. For every $N \in \NN$ we have 
\[
\text{tr}(\cC_T)  = \sum_{i \le N} \qv{\phi_i, \cC_T\phi_i} + \sum_{i > N} \qv{\phi_i, \cC_T\phi_i}.
\]

To bound the second sum on the right we note that
in view of (\ref{eq:covOp}) we have  $\cC_T^{-1} \ge  \cC_0^{-1}$.
Multiply this inequality by $\cC_0^{{1}/{2}}$ from both sides to obtain 
$I+\cC_0^{{1}/{2}}L_T^\circ \cC_0^{{1}/{2}}\geq I$ and then, noting
that $\cC_0^{{1}/{2}}L_T^\circ \cC_0^{{1}/{2}}$ is a bounded positive definite
symmetric operator, multiply the inequality with $( I+\cC_0^{{1}/{2}}L_T^\circ
\cC_0^{{1}/{2}})^{-{{1}/{2}}}$ on both sides to obtain 
$(I+\cC_0^{{1}/{2}} L_T^\circ \cC_0^{{1}/{2}})^{-1} \leq I$. Finally
multiply both sides by $\cC_0^\frac{1}{2}$ to arrive at $\cC_T \le \cC_0$. Naturally,
care has to be taken with the domain of definition of the unbounded operators
involved, but first performing the calculations for the Fourier basis functions 
$\phi_k$, one can pass to the limit, exploiting that
each multiplication by $\cC_0^\frac{1}{2}$ only adds compactness, see also
Exercise 8, p.243 of \cite{ConwayFuncAn} and the treatment in that chapter for
more details.

Hence, since $\lambda_i \sim i^{-2p}$, the second sum is bounded by a constant times  $N^{1-2p}$. 
By Cauchy-Schwarz the first sum is bounded by $\sum_{i \le N} \|\cC_T\phi_i\|_{L^2}$. 
To further bound this, we observe that 
\begin{align*}
\inf_{x \in \bbT} L^\circ_T(x; X) \|\cC_T\phi_i\|^2_{L^2} 
& \le \int_0^1 (\cC_T\phi_i(x))^2 L^\circ_T(x; X)\,dx\\
& \le \int_0^1 \cC_T\phi_i(x) (\cC_0^{-1} + L^\circ_T(\cdot; X))\cC_T\phi_i(x)\,dx\\
& = \int_0^1 \phi_i(x)\cC_T\phi_i(x) \,dx\\
& \le \|\phi_i\|_{L^2}\|\cC_T\phi_i\|_{L^2} = \|\cC_T\phi_i\|_{L^2}. 
\end{align*}
Dividing by $\|\cC_T\phi_i\|_{L^2}$ shows that 
$\|\cC_T\phi_i\|_{L^2} \le 1/\inf_{x \in \bbT} L^\circ_T(x; X)$ and hence, 
by the first statement of Theorem \ref{thm: local}, 
$\|\cC_T\phi_i\|_{L^2} = \OoP(1/T)$. 

Combining what we have we see that $\text{tr}(\cC_T) \le N \OoP(1/T) + N^{1-2p}$ 
for every $N \in \NN$. The choice $N \sim T^{1/(2p)}$ balances the two terms and shows that 
$\text{tr}(\cC_T) =\OoP(T^{(1-2p)/(2p)}) = \OoP(\eps^2_T)$.
\end{prf}

\begin{remarks}
\label{rem:dis}

It is clear from the proof of Theorem \ref{thm: rate}
that the posterior spread 
$\int \|\hat b_T-b_0\|^2_{L^2}\Pi(db\given X)$
is always of the order $T^{(1-2p)/(2p)}$, 
regardless of the smoothness of the true drift function $b_0$. 
Hence if the rate result of Theorem \ref{thm: mean} for the posterior mean can be improved, 
for instance the condition that $b_0 \in  \dot H^p(\bbT)$
can be relaxed to the assumption $b_0 \in \dot H^{p-1/2}(\bbT)$ (see the discussion in the concluding section), 
or the $\delta$ can be removed from the rate, then the result of Theorem \ref{thm: rate} 
for the full posterior automatically improves as well. 

We also note that the proof of Theorem \ref{thm: mean}
delivers convergence rates in other norms.
In particular it yields 
\[
\|\hat b_T - b_0\|_{H^p} = O_{\PP_{b_0}}(T^{\frac{1}{4p} +\delta})
\]
and hence, by interpolation (\cite{Robinson}, Lemma 3.27)
we have that the
error in the mean converges to zero as 
\[
\|\hat b_T - b_0\|_{H^s} = O_{\PP_{b_0}}(T^{\frac{1-2(p-s)}{4p} +\delta})
\]
for $0 \le s < p-1/2.$
\end{remarks}

\section{Proof of Theorem \ref{thm: local}}
\label{sec:proof}

\subsection{Semi-martingale versus diffusion local time}

Throughout this whole Section \ref{sec:proof}, the drift function $b \in \dot C(\bbT)$ 
is fixed and, contrary to our use in previous sections, we denote the underlying law 
by $\PP_x$ when the diffusion is started in $x$ and we sometimes shorten this to just 
$\PP$ when the diffusion is started in $0$.

The weak solution $X$ of the SDE (\ref{eq:diff}) is a regular diffusion on $\RR$ 
with scale function $s$ given by 
\[
s(x) = \int_{x_0}^x e^{-2\int_{y_0}^yb(z)\,dz}\,dy.
\]
We choose $x_0$ and $y_0$ such that $s(0) = 0$ and $s(1) = 1$. 
The speed measure $m$ has Lebesgue density $1/s'$. 
Since $b$ is $1$-periodic and mean-zero the function $s'$ is $1$-periodic as well. It follows that
$m$ is $1$-periodic and that $s$  satisfies 
\begin{equation}\label{eq: s}
s(x+k) = s(x) + k, 
\end{equation}
 for all $x \in \RR$ and $k \in \ZZ$.

The periodic local time $L^\circ$ was defined through the semi-martingale local time $L$ 
of the diffusion $X$, for which we have the occupation times formula (\ref{eq: occs}).
The diffusion $X$ also has continuous local time relative to its speed measure, 
the so-called diffusion local time of $X$. 
We denote this random field by $(\ell_t(x): t \ge 0, x \in \RR)$. 
It holds that $t \mapsto \ell_t(x)$ 
is continuous and for every $t \ge 0$ and bounded, measurable function $f$, 
\begin{equation}\label{eq: o}
\int_0^t f(X_u)\,du = \int_\RR f(x)\ell_t(x)\,m(dx)
\end{equation}
(see for instance \cite{Ito}). 
For this local time we define a periodic version $\ell^\circ$ as well, by setting
\[
\ell^\circ_t(x) = \sum_{k \in \ZZ}\ell_t(x + k). 
\]
The periodicity of $m$ then implies that for every $1$-periodic, bounded, measurable function $f$,
\begin{equation}\label{eq: occm}
\int_0^t f(X_u)\,du = \int_0^1 f(x)\ell^\circ_t(x)\,m(dx).
\end{equation}
Comparing this with (\ref{eq: occp1}) we see that we have the relation 
$s'(x)L^\circ_T(x; X) = \ell^\circ_T(x)$ for every $T \ge 0$ and $x \in [0,1]$. 
Now note that  $1/s'$ is up to a constant equal to the invariant density $\rho$
defined by (\ref{eq: rho}). Since $\rho$ is a probability density on $[0,1]$ and 
$1/s'$ is the density of the speed measure $m$, we have $m[0,1]\rho = 1/s'$ on $[0,1]$.  
Therefore,  statement (i) of Theorem \ref{thm: local} is equivalent to the statement that
\begin{equation}\label{eq: ulln}
\sup_{x \in \bbT} \Big|\frac{1}{t}\ell^\circ_t(x) - \frac1{m[0,1]}\Big| \to 0
\end{equation}
a.s.\ as $t \to \infty$, and statement (ii) is equivalent to the asymptotic tightness of 
\begin{equation}\label{eq: map}
x \mapsto \sqrt{t}\Big(\frac{1}{t}\ell^\circ_t(x) - \frac1{m[0,1]}\Big)
\end{equation}
 in $H^\alpha(\bbT)$ for every $\alpha \in [0,1/2)$. 
We will prove these statements in the subsequent subsections.

\subsection{A representation of the local time up to winding times}

%In this whole section the drift function $b$ is fixed and, contrary to our use in previous
%sections, we denote the underlying law by $\PP_x$ when the diffusion is started in $x$ and we
%sometimes shorten this to just $\PP$ when the diffusion is started in $0$. 
%If we do not write it explicitly, we work under the measure $\PP_0$, i.e.\ $X$ is started in $0$. 
We define a sequence of $\PP_0$-a.s.\ finite stopping times $\tau_0, \tau_1, \ldots$ by setting
$\tau_0 = 0$, $\tau_1$ is the first time $X$ exits $[-1, 1]$, $\tau_2$ is the first time 
after $\tau_1$ that $X$ exits $[X_{\tau_1} - 1, X_{\tau_1} +1]$, etc. 
(Note that if we define a process $Z$ on the complex unit circle by $Z_t = \exp(2i\pi X_t)$, 
then $\tau_k$ is the time that the process $Z$ completes its $k$th winding of the circle.)

The following theorem gives a representation for the periodic local time of $X$ up till the 
$n$th winding time. The representation involves a stochastic integral relative to $s(X)$. 
The process $s(X)$ is a diffusion in natural scale, hence a time-changed Brownian motion, and hence
a continuous local martingale.

\begin{theorem}\label{thm: rep}
For $x \in (0,1)$, 
\[
\frac1n \ell^\circ_{\tau_n}(x) - 1 = \frac1n\sum_{k=1}^{n} U_k(x), 
\]
where $U_1, \ldots, U_n$ are i.i.d.\ continuous random functions, distributed as 
\begin{align}
U(x) & = \ell_{\tau_1}(x) + \ell_{\tau_1}(x-1) - 1 \label{eq: u}\\
& = X_{\tau_1}(1-2s(x)) + 2\int_0^{\tau_1}\phi_x(X_u)\,ds(X_u) \label{eq: u2}, 
\end{align}
where $\phi_x = 1_{(x-1, \infty)} - 1_{(-\infty, x]}$.
\end{theorem}

\begin{prf}
For $k \in \NN$ we write $X^k = (X_{\tau_{k-1}+t}-X_{\tau_{k-1}}: t \ge 0)$
and $\tau^k_1 = \inf\{t: |X^k_t| = 1\}$.  By Lemma \ref{lem: markov} ahead, 
 the processes  $(X^k_t: t \in [0, \tau^k_1])$ are independent 
and have the same distribution as $(X_t: t \in [0, \tau_1])$. 
It follows  that for  $x \in (0,1)$, with $\ell^Z$ denoting the diffusion local time of 
the diffusion $Z$, 
\begin{equation}\label{eq: iid}
\frac1n \ell^\circ_{\tau_n}(x) - 1 = \frac1n\sum_{k=1}^{n} U_k(x), 
\end{equation}
where 
\[
U_k(x) = \ell^{X^k}_{\tau^k_1}(x) + \ell^{X^k}_{\tau^k_1}(x-1) - 1, 
\]
and the $U_k$ are independent copies of the random function
$U$ defined by (\ref{eq: u}).

Now let $Y = s(X)$. Then $Y$ is a regular diffusion in natural scale 
(i.e.\ the identity function is its scale function) and the speed measure $m^Y$ of $Y$ 
is related to the speed measure $m$ of $X$ by $m = m^Y \circ s$. 
%
%It follows from the periodicity of 
%$m$ and (\ref{eq: sinv}) that $m^Y$ is $1$-periodic as well.
It is easily seen that for $\ell^Y$ the local time of $Y$ relative to its speed measure $m^Y$, 
we have $\ell_t(x) = \ell^Y_t(s(x))$. 
For diffusions in natural scale, the diffusion local time coincides with 
the semi-martingale local time (see \cite{RogWil}, Section V.49). In particular, 
the Tanaka-M\'eyer formula holds: 
\begin{equation}\label{eq: tanaka1}
\ell^Y_t(x) = |Y_t - x| - |x| - \int_0^t \text{sign}(Y_u- x)\,dY_u
\end{equation}
under $\PP_0$. In view of (\ref{eq: s}) $\tau_1$ is also the first time that $Y$ exits $[-1,1]$, 
so we have that $X_{\tau_1} = Y_{\tau_1}$.  
Using also the fact that the scale function $s$ is strictly increasing, we obtain
\begin{equation}\label{eq: tanaka}
\ell_{\tau_1}(x) = |X_{\tau_1} - s(x)| - |s(x)| - \int_0^{\tau_1} \text{sign}(X_u- x)\,ds(X_u).
\end{equation}
Together with (\ref{eq: s}) this implies that (\ref{eq: u2}) holds. 
\end{prf}

The proof of the theorem uses the following lemma, which implies that $X$  ``starts afresh''
after every winding time $\tau_k$. Let $(\FF_t: t \ge 0)$ denote the natural 
filtration of the process $X$. 

\begin{lemma}\label{lem: markov}
For every $\PP_0$-a.s.\ finite  stopping time $\tau$ such that $X_\tau \in \ZZ$ a.s., 
it holds that the process $(X_{\tau + t} - X_\tau: t \ge 0)$ is independent of $\FF_\tau$ and 
has the same law as $X$ under $\PP_0$. 
\end{lemma}

\begin{prf}
Fix a measurable subset $C \subset C[0,\infty)$. By the strong Markov property we have 
\[
\PP_0(X_{\tau + \cdot} - X_\tau \in C \given \FF_\tau) = f(X_\tau)
\]
a.s., where $f(x) = \PP_x(X - X_0 \in C)$.  The periodicity of the drift function implies that 
for every $k \in \ZZ$, 
$f(k)  = \PP_k(X-k \in C) = \PP_0(X \in C)$. 
Hence we have 
\[
\PP_0(X_{\tau + \cdot} - X_\tau \in C \given \FF_\tau) = \PP_0(X \in C), 
\]
a.s., which completes the proof.
\end{prf}

Since we will be interested in the local time up till a deterministic time $t$, 
it is necessary to deal with the time interval between $t$ and the previous or next winding time. 
The following lemma will be used for that. For $t \ge  0$, let the $\ZZ_+$-valued 
random variable $n_t$ be such that $\tau_{n_t}$ is the last winding time less or equal to  $t$, 
so $\tau_{n_t} \le t < \tau_{n_t+1}$. 
%For $a \in \RR$, let $\PP_a$ be the 
%law under which the process $X$ starts in $a$. 
%WE SAID THIS BEFORE

\begin{lemma}\label{lem: rest}
For all $t \ge 0$ and Borel sets $B \subset C[0,1]$, 
\[
\PP_0(\ell^\circ_{\tau_{n_t +1}} - \ell_{t}^\circ \in B) = 
\bbE_0 \PP_{X_t-X_{n_t}}(\ell^\circ_{\tau_1} \in B).
\]
\end{lemma}

\begin{prf}
We split up the event of interest according to the position of $X$ at time $\tau_{n_t}$. 
For $k \in \ZZ$ we have 
\begin{align*}
\PP_0(\ell^\circ_{\tau_{n_t +1}} - \ell_{t}^\circ \in B; X_{\tau_{n_t}} = k) = 
\PP_0(\ell^\circ_{\sigma_{t, k}} - \ell_{t}^\circ \in B; X_{\tau_{n_t}} = k), 
\end{align*}
where $\sigma_{t, k} = \inf\{s >t: |X_s - k| \ge  1\}$.
Let $(\mathcal{F}_s: s \ge 0)$ be the natural filtration of the process $X$.
Since $X_{\tau_{n_t}}$ is $\FF_t$-measurable, conditioning on $\FF_t$ gives  
\[
\PP_0(\ell^\circ_{\tau_{n_t +1}} - \ell_{t}^\circ \in B; X_{\tau_{n_t}} = k) 
= \bbE_0 1_{\{X_{\tau_{n_t}} = k\}}
    \PP_0(\ell^\circ_{\sigma_{t, k}} - \ell_{t}^\circ \in B \,|\,\FF_t).
\]
By the Markov property, the conditional probability equals
$\PP_{X_t}(\ell^\circ_{\sigma_{0, k}} \in B)$. 
By the periodicity of the drift function, this is equal to 
$\PP_{X_t-k}(\ell^\circ_{\sigma_{0, 0}} \in B)$. Since $\sigma_{0,0} = \tau_1$, we obtain
\[
\PP_0(\ell^\circ_{\tau_{n_t +1}} - \ell_{t}^\circ \in B; X_{\tau_{n_t}} = k) = 
\bbE_0 1_{\{X_{\tau_{n_t}} = k\}}\PP_{X_t-k}(\ell^\circ_{\tau_1} \in B).
\]
Summation over $k$ completes the proof.
\end{prf}

\subsection{Proof of statement (i)  of Theorem \ref{thm: local}}

In this subsection we prove that (\ref{eq: ulln}) holds a.s.\ for $T \to \infty$, which is 
equivalent to statement (i)  of Theorem \ref{thm: local}.

According to Theorem \ref{thm: rep} we have 
\[
\frac1n \ell^\circ_{\tau_n}(x) - 1 = \frac1n\sum_{k=1}^{n} U_k(x), 
\]
where the $U_k$ are independent copies of the continuous random function on $[0,1]$ 
given by (\ref{eq: u}). Now 
$\bbE \|U\|_\infty \le 1+2\bbE \sup_{|x| \le 1} \ell_{\tau_1}(x)$. To bound the expectation, 
we again use the fact that $\ell_{\tau_1}(x) = \ell^{Y}_{\tau_1}(s(x))$, for $Y = s(X)$. 
Relation (\ref{eq: s}) implies that $\sup_{|x| \le 1} \ell_{\tau_1}(x) = 
\sup_{|x| \le 1} \ell^Y_{\tau_1}(x)$. 
Applying the  BDG-type inequality for local times to the stopped continuous 
local martingale $Y^{\tau_1}$ (see \cite{Revuz}, Theorem XI.(2.4))
we then see that for some constant $C > 0$, 
\[
\bbE \|U\|_\infty \le 1+2\bbE \sup_{|x| \le 1} \ell^Y_{\tau_1}(x) \le 1 + C \bbE 
\sup_{t \le \tau_1}|Y_t| < \infty. 
\]
Since by (\ref{eq: s}) it holds that $X_{\tau_1} = \pm 1$ with equal probability, 
it easily derived from (\ref{eq: u2}) that $\EE U(x) = 0$. 
By the Banach space version of Kolmogorov's law of large numbers (see \cite{Led91}, Corollary 7.10), 
it follows that 
\begin{equation}\label{eq: llntau}
\sup_{x \in [0,1]} \Big| \frac1n \ell^\circ_{\tau_n}(x) - 1\Big| \to 0
\end{equation}
a.s.. 

The random variables $\tau_1, \tau_2- \tau_1, \tau_3-\tau_2, \ldots$ are i.i.d., 
so by the law of large numbers, $\tau_n/n \to \EE\tau_1$ a.s.. 
Applying relation (\ref{eq: o}) with $t = \tau_1$ and $f \equiv 1$ we see that 
\[
\bbE \tau_1 = \int_{-1}^1 \EE\ell_{\tau_1}(x)\,m(dx). 
\] 
Since $X_{\tau_1} = \pm 1$ with equal probability, (\ref{eq: tanaka}) implies 
that $\EE\ell_{\tau_1}(x) = 1-|s(x)|$. Using (\ref{eq: s}) and the periodicity of $m$, 
it follows that 
\begin{align*}
\EE\tau_1 & = \int_{-1}^0(1+s(x))\,m(dx) + \int_0^1(1-s(x))\,m(dx) \\
& = \int_{0}^1(1+s(x-1))\,m(dx) + \int_0^1(1-s(x))\,m(dx) = m[0,1]. 
\end{align*}
Combining (\ref{eq: llntau}) with the fact that $\tau_n/n \to m[0,1]$ a.s., we find that 
\begin{equation}\label{eq: llntau2}
\sup_{x \in [0,1]} \Big| \frac1{\tau_n} \ell^\circ_{\tau_n}(x) - \frac1{m[0,1]}\Big| \to 0
\end{equation}
a.s.. 

Now let $n_t$ be defined as before Lemma \ref{lem: rest}, so that $\tau_{n_t} \le t < \tau_{n_t+1}$.
Then as $t \to \infty$ it holds that $n_t \to \infty$ and hence 
${\tau_{n_t}}/{n_t} \to m[0,1]$ a.s.\ and ${\tau_{n_t+1}}/{n_t} \to m[0,1]$ a.s.. 
It follows that $n_t / t \to 1/m[0,1]$ a.s., and therefore also $\tau_{n_t}/t \to 1$ a.s.. 
We can write 
\[
\frac1t\ell^\circ_t(x) = \frac{\tau_{n_t}}{t}\frac1{\tau_{n_t}}\ell^\circ_{\tau_{n_t}}(x) + 
\frac{1}{t}(\ell^\circ_{t}(x)-\ell^\circ_{\tau_{n_t}}(x)).
\]
Relation (\ref{eq: llntau2}) shows that a.s., the first term on the right converges 
uniformly to $1/m[0,1]$. The second term is non-negative and bounded by 
\[
\frac1t(\ell^\circ_{\tau_{n_t} + 1}(x)-\ell^\circ_{\tau_{n_t}}(x)) = 
\frac{\tau_{n_t+1}}{t}\frac1{\tau_{n_t+1}}
\ell^\circ_{\tau_{n_t} + 1}(x)
- \frac{\tau_{n_t}}{t}\frac1{\tau_{n_t}}\ell^\circ_{\tau_{n_t} }(x), 
\]
which converges  uniformly to $0$ by the preceding. 
This completes the proof of (\ref{eq: ulln}) and hence of statement (i) of Theorem \ref{thm: local}.

\subsection{Proof of statement (ii)  of Theorem \ref{thm: local}}
In this subsection we prove that the random maps (\ref{eq: map}) are asymptotically 
tight in the space  in $H^\alpha(\bbT)$ for every $\alpha \in [0,1/2)$, which is 
equivalent to statement (ii)  of Theorem \ref{thm: local}.
It is most convenient and of course not restrictive to work with the complex Sobolev spaces. 
Let $e_k(x) = \exp(i2k\pi x)$, $k \in \ZZ$,  be the standard complex exponential 
basis of $L^2[0,1]$. For $\alpha \ge 0$, define the associated Sobolev  space
\[
H^\alpha[0,1] = \Big\{f \in L^2[0,1]:   
      \|f\|_{H^\alpha}^2 = \sum |k|^{2\alpha}|\qv{f, e_k}|^2 < \infty\Big\},
\]
where $\qv{f,g} = \int_0^1 f(x)\bar g(x)\,dx$ is the usual inner product on $L^2[0,1]$.

By the representation  of the local time given by Theorem \ref{thm: rep} and the 
central limit theorem for Hilbert space-valued random elements (e.g.\ \cite{Led91}, Corollary 10.9), we have that 
\begin{equation}\label{eq: weak}
\sqrt{n}\Big(\frac1n \ell_{\tau_n}^\circ - 1\Big)
\end{equation}
converges weakly in $H^\alpha[0,1]$ if 
\begin{enumerate}
\item
$\EE \|U\|^2_{H^\alpha} < \infty$.
\item
$\EE U = 0$ (where the expectation is to be interpreted as a Pettis integral),
\end{enumerate}
We will show that these conditions hold if (and only if) $\alpha < 1/2$. 
Slightly abusing notation, denote the two functions on the right of  (\ref{eq: u2})
by $U_1$ and $U_2$. We will show that conditions 1--2 hold for 
$U_1$ and $U_2$ separately.

To show that the  
conditions hold for $U_1$, recall that   $X_{\tau_1} \pm 1$ with equal probability. 
Hence, $\EE U_1 = 0$ and $\EE \|U_1\|^2_{H^\alpha} = \|1-2s\|^2_{H^\alpha} < \infty$.

As for $U_2$, using (\ref{eq: u2}) and the stochastic Fubini theorem  it is readily checked that
\[
\qv{ U_2, e_k} = 
2\int_0^{\tau_1}c_k(X_u)\,ds(X_u), 
\]
where
\begin{align*}
c_k(x) &= 
\begin{cases}
\displaystyle\frac{1-e^{i2k\pi}}{i2k\pi} & \text{if $x +1 \le 0$},\\[1.5ex]
\displaystyle\frac{e^{i2k(x+1)\pi} - e^{i2k\pi}}{i2k\pi} & \text{if $x \le 0 \le x +1 \le 1$},
\\[1.5ex] 
\displaystyle\frac{e^{i2k\pi x} - 1}{i2k\pi} & \text{if $0 \le x \le 1 \le x +1 $},\\[1.5ex] 
\displaystyle\frac{e^{i2k\pi} - 1}{i2k\pi} & \text{if $x \ge 1 $}. 
\end{cases}
\end{align*}
To show that condition 1.\ holds for $U_2$, note that for $u \le \tau_1$ it holds that $|X_u| \le 1$. 
It is straightforward to see that for $|x| \le 1$, we have 
$|c_k( x)| \le C(1+|k|)^{-1}$ for some $C > 0$. Therefore, by the It\^o isometry, 
\begin{align*}
\EE\sum|k|^{2\alpha}\Big|\int_0^{\tau_1}c_k(X_u)\,ds(X_u)\Big|^2
& = \sum |k|^{2\alpha} \EE\int_0^{\tau_1}\Big|c_k(X_u)\Big|^2\,d\qv{s(X)}_u\\
& \le C^2\EE\qv{s(X)}_{\tau_1}\sum \frac{|k|^{2\alpha}}{(1+|k|)^2}.
\end{align*}
The sum on the right is finite if $\alpha \in [0,1/2)$.  
For the diffusion $Y=s(X)$ the diffusion local time coincides with the semi-martingale local time, 
hence
\[
\EE \qv{s(X)}_{\tau_1} = \EE \int \ell^Y_{\tau_1}(x)\,dx = 
\int_{-1}^1 \EE \ell^Y_{\tau_1}(x)\,dx.
\]
The Tanaka-M\'eyer formula and optional stopping imply that for $|x| \le 1$, 
\[
\EE \ell^Y_{\tau_1}(x) = \EE |Y_{\tau_1} - x| - |x|  = 
1- |x|. 
\]
Hence, $U_2$ satisfies
 condition 1.
Finally, note that to show that $\EE U_2 = 0$, it suffices to show that $\EE U_2(x) = 0 $ for 
every fixed $x \in (0,1)$. But this follows readily from (\ref{eq: u}) again, by optional stopping. 
So indeed the random maps (\ref{eq: weak}) converge in $H^\alpha[0,1]$ 
for every $\alpha \in [0,1/2)$.

To complete the proof we consider the decomposition
\begin{align*}
& \sqrt{t}\Big(\frac{\ell^\circ_{t}}{t}-\frac1{m[0,1]}\Big)\\
&  = \sqrt{n_t+1}\Big(\frac{\ell_{\tau_{n_t+1}}^\circ }{n_t+1} - 1\Big)\sqrt{\frac{n_t+1}{t}}
+ \sqrt{t}\Big(\frac{n_t+1}{t}- \frac1{m[0,1]}\Big)
- \frac{\ell^\circ_{\tau_{n_t+1}}- \ell^\circ_t}{\sqrt t}.
\end{align*}
Since $n_t/t \to 1/m[0,1]$ a.s., the tightness of the maps (\ref{eq: weak}) 
implies that the first term is asymptotically tight. By the central limit theorem, 
$\sqrt{n}(\tau_n/n- m[0,1])$ converges in distribution. Together with the inequality 
$\tau_{n_t} \le t < \tau_{n_t+1}$ and the delta method this implies that the second term 
is asymptotically tight as well. For the last term, note that 
by Lemma \ref{lem: rest} we have, for $M > 0$,  
\[
\PP_0\Big(\Big\|\frac{\ell^\circ_{\tau_{n_t+1}}- \ell^\circ_t}{\sqrt t}\Big\|_{H^\alpha} > M\Big)
\le \sup_{|a| \le 1} \PP_a(\|\ell_{\tau_1}\|_{H^\alpha} > 
    M\sqrt{t}) \le \frac1{M^2 t}\sup_{|a| \le 1}\EE_a \|\ell_{\tau_1}\|^2_{H^\alpha}.
\]
Similar considerations as used to show that condition 1.\ above holds for $U_2$ 
show that the supremum over $a$ on the right-hand side is bounded.
We conclude that the last term in the decomposition is $o_P(1)$. This completes the proof.

\section{Concluding remarks}

We have obtained the posterior contraction rate $T^{-(p-1/2)/(2p)}$ for our nonparametric 
Bayes procedure. We remarked that the regularity of the prior is essentially $p-1/2$
(Lemma \ref{lem: prior}) and assumed that the true drift  $b_0$ has Sobolev regularity of order $p$.
Although lower bounds for the rate of convergence in the exact model under study do not appear 
to be known, comparison with similar models suggests that the optimal rate for estimating a 
drift function $b_0$ that is $\beta$-regular (in Sobolev sense) may be $T^{-\beta/(1+2\beta)}$ in our setting 
(in a minimax sense over Sobolev balls for instance, cf.\ e.g.\ \cite{Kutoyants04}, \cite{Tsybakov}
for similar results). 
The general message from the Gaussian process prior literature is that this optimal rate 
is typically attained if the ``regularity'' $\alpha$ of the prior matches the regularity $\beta$ 
of the function that is being estimated (see \cite{vZRate}).
Since the regularity of the prior we employ 
in this paper is essentially $\alpha = p-1/2$, this suggests that in principle, 
it should be possible to relax our assumption 
that $b_0$ is $p$-regular 
to the assumption that  
$b_0$ is $(p-1/2)$-regular, while still maintaining the same rate $T^{-(p-1/2)/(2p)}$.
It is however not clear 
whether this can be achieved by adapting the proof we give in this paper. 
The method of proof is adapted from \cite{ALS12} where it
is used to study linear inverse problems in the small noise
limit. In that context the proof gives sharp rates in some
parameter regimes, but not in others.

There are a number of future directions that this work
could be taken in. First of all, alternative technical approaches could be explored 
to derive sharp convergence rates.
One  approach could be to use the representation of the posterior mean 
as a minimizer of some stochastic objective functional (cf.\ Subsection \ref{ssec: pen}) and use 
empirical process-type techniques to study its asymptotic properties. This however requires 
technical tools (e.g.\ uniform limit theorems, maximal inequalities) 
that are presently not available in this setting of periodic diffusions. 
Alternatively, sharp rates may result from a general rate of convergence theory for posteriors 
in the spirit of \cite{MVZ}, if that could be developed for 
this class of models.
Secondly, motivated by practical considerations,
it will also be interesting to determine whether useful adaptive 
procedures can be constructed by choosing the hyper-parameters 
$p, \eta$ and $\kappa$ in a data-driven way, 
for instance by hierarchical Bayes or empirical Bayes procedures.
There is recent computational work in this direction, cf.\ \cite{Moritz}, 
but no theoretical results are presently available.
A third future direction concerns extension of
the ideas in this paper
to diffusions in more than one dimension. The
local time is, then, a much more singular object and 
developing an analysis of posterior consistency will
present new challenges.

\subsection*{Acknowledgements}

The research of AMS is funded by the EPSRC (UK) and by the ERC. 
The research of JHvZ is  supported by the Netherlands 
Organization for Scientific Research NWO.
The authors are grateful to Sergios Agapiou for helpful comments.

\end{document}